# Diffractive Deep-Inelastic Scattering[1]

Stefan Tapprogge

Institut für Hochenergiephysik, Universität Heidelberg
Schröderstr. 90
D 69120 Heidelberg, Germany

for the H1 Collaboration

**Abstract**

New results on diffractive deep-inelastic $ep$ scattering at HERA are presented using data taken in 1994 with the H1 detector. The cross section for diffractive deep-inelastic scattering is measured in terms of a diffractive structure function $F_2^{D(3)}(\beta, Q^2, x_{I\!P})$ over an extended kinematic range. The dependence of $F_2^{D(3)}$ on $x_{I\!P}$ is found not to depend on $Q^2$, but to depend on $\beta$. Therefore the $x_{I\!P}$ dependence no longer factorizes. The $Q^2$ and $\beta$ dependence of $F_2^{D(3)}$ is analyzed after an integration over the dependence on $x_{I\!P}$. For fixed $\beta$ a clear rise with $\log Q^2$ is observed, persisting up to high values of $\beta$. In terms of the Altarelli-Parisi (DGLAP) QCD evolution equations, these scaling violations give clear indications for a gluon dominated process. Subsequently an attempt is made to quantify the parton content of the diffractive exchange using the DGLAP evolution. At the starting scale a "leading" gluon distribution is found which contributes about 80% of the momentum in the diffractive exchange. Measurements of the hadronic final state (energy flow and production of $D^*$ mesons) are found to be consistent with the predictions of a model of deep-inelastic electron pomeron scattering using the information on the parton content obtained.

---

# 1 Introduction

The data taken in 1992 at HERA led to the first observation of deep-inelastic scattering (DIS) events with a large gap in rapidity around the proton remnant direction[1, 2]. The data from 1993 allowed to measure the cross section for this contribution to DIS, quantified in terms of a diffractive structure function[3, 4]. It has been demonstrated that this process is dominantly of a diffractive nature. No substantial $Q^2$ dependence was observed, indicating approximate scale invariance and thus a pointlike nature of the process. First indications for a large contribution from gluonic processes were obtained[5].

The increase in statistics ($\int \mathcal{L} dt \approx 2$ pb$^{-1}$) in the 1994 data and the access to lower values of $Q^2$ during a dedicated running period allow further probing of diffractive deep-inelastic scattering. This contribution presents new H1 results on the diffractive structure function and on the hadronic final state in diffractive DIS.

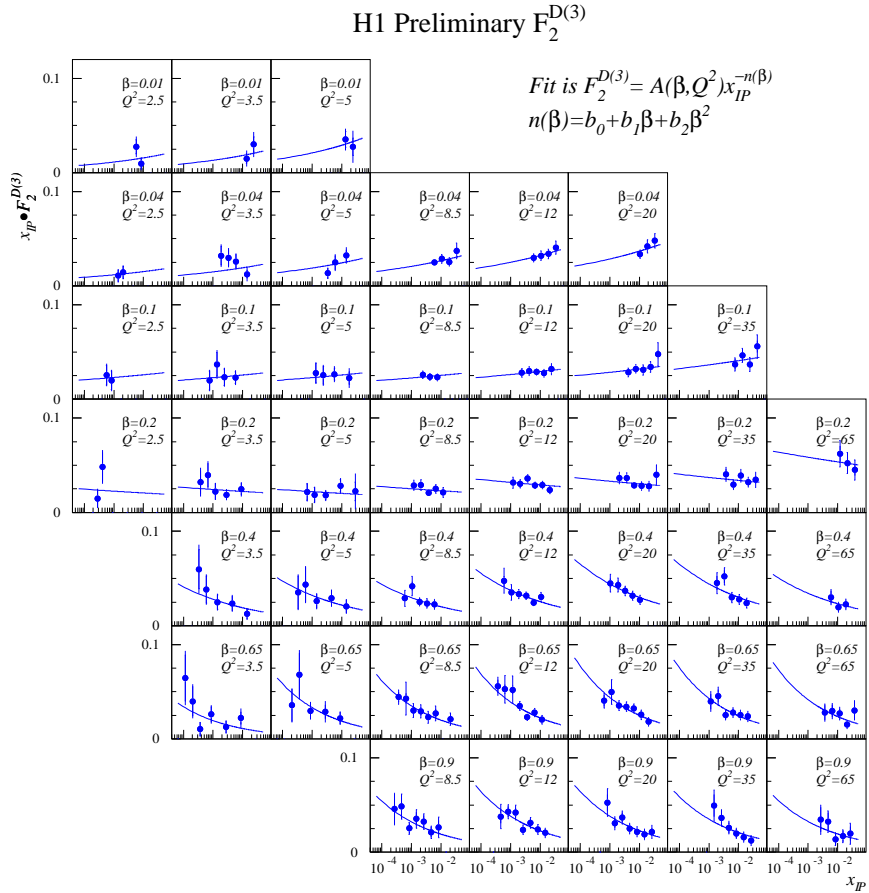

Figure 1: $x_{I\!P} \cdot F_2^{D(3)}(\beta, Q^2, x_{I\!P})$ together with a fit of the form $F_2^{D(3)} \propto A(\beta, Q^2) \cdot x_{I\!P}^{-n(\beta)}$

# 2 Cross Section for Diffractive Deep-Inelastic Scattering

The process under investigation is $ep \to e'XY$, where the systems $X$ and $Y$ are separated by the largest gap in rapidity in the hadronic final state. This forces system $Y$ kinematically to be a "leading" system which in most cases follows the proton beam direction ("forward" direction in the H1 coordinate system). System $X$ is measured in the central detector. The experimental selection for this process demands no activity in the forward components of the H1 detector giving access to the region $3.4 < \eta < 7.0$ in pseudo-rapidity $\eta$ $(= -\log\tan(\theta/2))$. Defining $Q^2 = -q^2$ and $W^2 = (q + P)^2$ where $q$ ($P$) is the four momentum of the photon (proton) and $M_Y$ ($M_X$) the invariant mass of system $Y$ ($X$), the process $ep \to e'XY$ can be studied for $M_Y < 1.6$ GeV/$c^2$ and $x_{I\!P} = (Q^2 + M_X^2)/(Q^2 + W^2) < 0.05$. Since $t = (P - P')^2$, where $P'$ is the four momentum of $Y$, is presently not measured, the data are used to determine the cross section (integrated over $t$ with $|t| < 7$ GeV$^2/c^2$)

$$\frac{d^3\sigma^D_{ep \to eXY}}{dx_{I\!P} \, d\beta \, dQ^2} = \frac{4\pi\alpha^2}{\beta Q^4}(1 - y + \frac{y^2}{2})F_2^{D(3)}(\beta, Q^2, x_{I\!P})$$

in the kinematic range $2.5 < Q^2 < 65$ GeV$^2/c^2$, $0.01 < \beta < 0.9$ and $0.0001 < x_{I\!P} < 0.05$. Here $\beta = Q^2/(Q^2 + M_X^2)$ and $F_2^{D(3)}(\beta, Q^2, x_{I\!P})$ is called the diffractive structure function. $F_2^{D(3)}$ is found to decrease in a power-law-like behaviour with increasing $x_{I\!P}$. Therefore in fig. 1 $x_{I\!P} \cdot F_2^{D(3)}(\beta, Q^2, x_{I\!P})$ is shown as a function of $x_{I\!P}$ in bins of $\beta$ and $Q^2$. Also shown is a fit $F_2^{D(3)} \propto A(\beta, Q^2) \cdot x_{I\!P}^{-n(\beta)}$ to the data, where $n(\beta) = b_0 + b_1\beta + b_2\beta^2$. It is clearly visible that there is no universal dependence of $F_2^{D(3)}$ on $x_{I\!P}$ for all $\beta$ and $Q^2$. The value obtained for $n$ decreases with decreasing $\beta$, whereas no dependence on $Q^2$ is observed. The 1994 data clearly show that the $x_{I\!P}$ dependence of $F_2^{D(3)}$ does no longer factorise from the dependence on $\beta$.

To investigate the partonic structure of the exchange in the diffractive process, the $x_{I\!P}$ dependence of $F_2^{D(3)}$ is averaged by an integration to define

$$\tilde{F}_2^D(\beta, Q^2) = \int_{x_{I\!P_L}=0.0003}^{x_{I\!P_H}=0.05} F_2^{D(3)}(\beta, Q^2, x_{I\!P}) dx_{I\!P}.$$

$\tilde{F}_2^D(\beta, Q^2)$ amounts to the structure function of the diffractive exchange averaged over $t$ and $x_{I\!P}$.

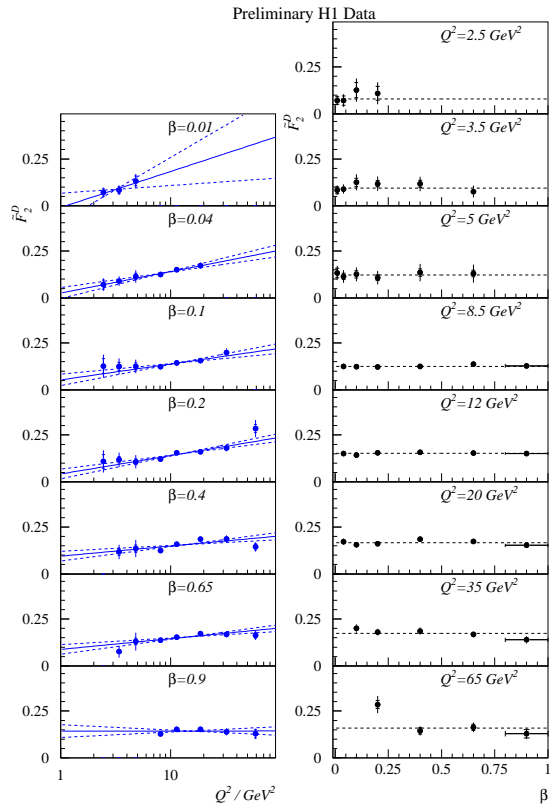

Figure 2: $\tilde{F}_2^D(\beta, Q^2)$ as a function of $Q^2$ for different $\beta$ values (left part) and as a function of $\beta$ for different $Q^2$ values (right part).

The left part of fig. 2 shows $\tilde{F}_2^D(\beta, Q^2)$ as a function of $Q^2$ for fixed $\beta$ and the right part shows $\tilde{F}_2^D(\beta, Q^2)$ as a function of $\beta$ for fixed $Q^2$. The $Q^2$ dependence exhibits a rise of $\tilde{F}_2^D$ with $\log Q^2$ which persists to high values of $\beta$, whereas the $\beta$ dependence for fixed $Q^2$ is flat. Interpreting the $Q^2$ dependence in terms of the Altarelli-Parisi (DGLAP) QCD evolution equations, these scaling violations constitute clear evidence for a gluon dominated process. This is supported by a QCD analysis of $\tilde{F}_2^D$. At a starting scale $Q_0^2 = 2.5$ GeV$^2/c^2$ singlet quark densities $q(x_{i/I\!P})$ are parametrized as $x_{i/I\!P} \cdot q(x_{i/I\!P}) = A x_{i/I\!P}^B (1 - x_{i/I\!P})^C$, where $x_{i/I\!P}$ is the fractional momentum. The same ansatz is used for the gluon density, but with different parameters $A$, $B$ and $C$. These parton densities are then evolved according to the DGLAP equations to higher values of $Q^2$. The contribution from charm quarks is taken into account via the boson–gluon fusion process following the same procedure as in[6]. The parton densities obtained from a fit to $\tilde{F}_2^D$ have more than 80% of the momentum carried by gluons for the $Q^2$ range investigated. The gluon density obtained from this fit shows a "leading" behaviour at $Q^2 \approx 5$ GeV$^2/c^2$, i.e. most gluons have a fractional momentum $x_{i/I\!P} > 0.9$ tending towards a singularity at $x_{i/I\!P} \approx 1$. This "leading" gluon behaviour resembles the expectation of the model by Buchmüller and Hebecker[7].

## 3 Measurements of the Hadronic Final State

Using the parton densities obtained from the DGLAP QCD analysis above and the RAPGAP model[8] in which deep-inelastic diffraction is understood in terms of deep-inelastic electron-pomeron $(eI\!P)$ scattering[9], predictions can be made for the hadronic final state and comparison made with observation.

For a gluon dominated process, more energy flow is expected in the central region $(\eta^* \approx 0)$ of the $\gamma^* I\!P$ centre-of-mass system compared to a quark dominated process. Fig. 3 shows the measured energy flow in three bins of $M_X$ compared to the prediction of the RAPGAP model using the parton densities obtained above. The data are described well for two different approaches to the modelling of the QCD cascade in the final state. Not shown is the prediction of the RAPGAP model with unphysical parton densities (only quarks, no gluons at all), which gives less energy flow in the central region $\eta^* \approx 0$ than in the data.

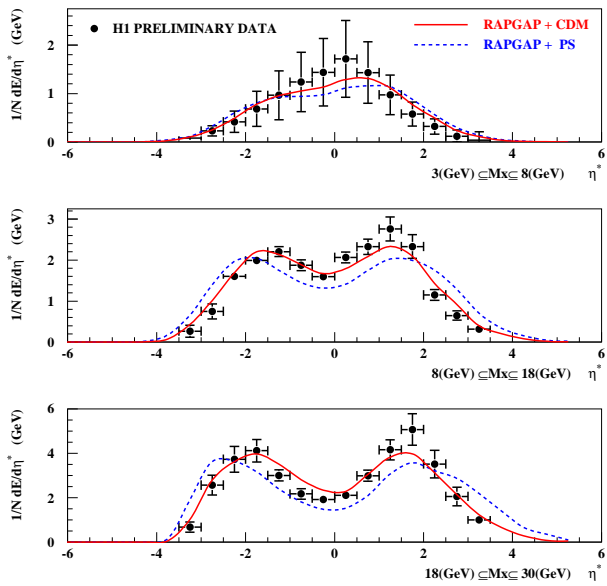

Figure 3: Energy flow in the $\gamma^* I\!\!P$ CMS – fully corrected for detector effects – compared to the RAPGAP model using the parton densities obtained from $\tilde{F}_2^D(\beta, Q^2)$ in which gluons dominate; shown are two different approaches to the modelling of QCD cascades (CDM = Colour Dipole Model, PS = Parton Shower in the "leading log" approximation). Errors shown are statistical and systematic errors added in quadrature

From the dominant gluon contribution, the production of charm quarks via the boson–gluon fusion process is expected.

$D^*$ meson production and decay has been searched for in the diffractive data by looking for the decay $D^{*\pm} \to D^0(\bar{D}^0)\pi_{slow}^{\pm} \to K^{\mp}\pi^{\pm}\pi_{slow}^{\pm}$. A signal for $D^*$ mesons is best seen in the difference of invariant masses $\Delta M = M_{K\pi\pi_{slow}} - M_{K\pi}$. Here the resolution is dominated by the measurement of the slow pion.

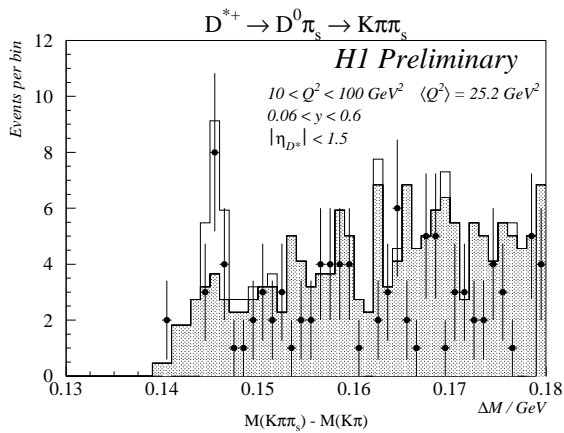

Figure 4: Mass difference between $K\pi\pi_{slow}$ and $K\pi$ demanding $|M_{K\pi} - M_{D^0}| < 80$ MeV/$c^2$. The histogram is the prediction of the RAPGAP model using the parton densities obtained from the QCD analysis; the shaded histogram shows those events of the model without a $D^*$ meson

Fig. 4 shows this mass difference for those $(K,\pi)$ combinations close to the $D^0$ mass ($|M_{K\pi} - M_{D^0}| < 80$ MeV/$c^2$). The $D^*$ yield in the data is found to be compatible with the prediction by the RAPGAP model (unshaded histogram) using the parton densities obtained.

These two measurements of the final state are consistent with the expectations of a model for deep-inelastic $eI\!\!P$ scattering where the parton densities for the pomeron $I\!\!P$ have been obtained from a QCD analysis of the inclusive diffractive structure function measurement.

## 4 Conclusions

Diffractive deep-inelastic $ep$ scattering at HERA has been investigated using the data taken in 1994 with the H1 detector. The diffractive structure function $F_2^{D(3)}(\beta, Q^2, x_{I\!\!P})$ was found to have no universal dependence on $x_{I\!\!P}$ for all $\beta$. The unambiguous observation of scaling violations in the $Q^2$ dependence for fixed $\beta$ (rising with $\log Q^2$ up to high $\beta$ values) indicated in the framework of the Altarelli-Parisi QCD evolution equations a gluon dominated process. A model for deep-inelastic $eI\!\!P$ scattering was found to concur with the measured hadronic final state (energy flow and $D^*$ meson production) when using parton densities obtained from a DGLAP QCD analysis of $F_2^{D(3)}$ which have a dominant "leading" gluon component.